\documentstyle[prl,aps,amsmath,amssymb,twocolumn]{revtex}

\begin{document}

\title{Zero Energy Solutions 
and Vortices 
in Schr\"{o}dinger Equations 
}

\author{Tsunehiro Kobayashi$^*$
and Toshiki Shimbori$^\dag$}

\address{$^*$Department of General Education 
for the Hearing Impaired, 
Tsukuba College of Technology, 
Ibaraki 305-0005, Japan \\ 
$^\dag$Institute of Physics, Unversity of Tsukuba, 
Ibaraki 305-8571, Japan}

\twocolumn[
\widetext
\begin{@twocolumnfalse}

\maketitle



\begin{abstract}
Two-dimensional Schr$\ddot {\rm o}$dinger equations 
with rotationally symmetric potentials 
$(V_a(\rho )= -a^2g_a \rho ^{2(a-1)}  {\rm with}\ \rho=\sqrt{x^2+y^2}
\ {\rm and}\ a\not=0)$ 
are shown to have zero energy states contained in conjugate spaces 
of Gel'fand triplets.  
For the zero energy eigenvalue the equations  for all $a$ 
are reduced to the same equation representing 
two-dimensional free motions 
in the constant potential $V_a=-g_a$ 
in terms of the conformal mappings of 
$\zeta_a=z^a$ with $z=x+iy$. 
Namely, the zero energy eigenstates are described by 
the plane waves with the fixed wave numbers $k_a=\sqrt{mg_a}/(h/2\pi)$ 
in the mapped spaces. 
All the zero energy states are infinitely degenerate as same as 
the case of 
the parabolic potential barrier (PPB) shown in ref. 8. 
Following hydrodynamical arguments, we see that 
such states describe stationary flows round the origin, 
which are represented by the complex velocity potentials $W=p_a z^a$, 
( $p_a$ being a complex number) 
and their linear combinations create almost 
arbitrary vortex patterns. 
Examples of the vortex patterns in constant potntials and PPB 
are presented. 
We see that any vortices cannot be created in the linear combinations 
of the plane-wave solutions 
and the degenerate states which are not described by plane-waves 
play essential roles to create vortices. 
In the extension to three-dimensional problems with potentials 
being separable into $2+1$ dimensions 
we show that the states in three dimensions 
have the same structure as the two-dimensional states 
with the zero energy but they 
can generally have non-zero total energies. 

\noindent 
PACS numbers: 03.65.-w, 03.50.-z

\end{abstract}
\end{@twocolumnfalse}
]
\narrowtext

\section{Introduction} \label{sect.1.0}
 
It is known that scattering states and unstable states like resonances 
are generally described by states in conjugate spaces 
of Gel'fand triplets (GT)~\cite{bohm}. 
An example of such states for the parabolic potential barrier (PPB) 
$V=-m\gamma^2 x^2/2$ 
in one dimension has 
been studied by many authors [2-7]. 
It has been shown that the one-dimensional PPB has pure imaginary 
energy eigenvalues $\mp i(n+1/2)\hbar \gamma$ with 
$n=0,1,2,\cdots$, and the eigenfunctions are generalized functions 
in the conjugate space ${\cal S}({\mathbb{R}})^\times $ of 
GT described by 
${\cal S}({\mathbb{R}}) \subset {\cal L}^2({\mathbb{R}})\subset 
{\cal S}({\mathbb{R}})^\times  $, 
where ${\cal S}({\mathbb{R}}) $ and ${\cal L}^2({\mathbb{R}})$, respectively, stand for 
a Schwartz space and a Lebesgue space~\cite{sk,s2}. 
In general the energy eigenvalues ${\cal E}$ of the 
conjugate spaces 
in GT are expressed by pairs of complex conjugates such that 
${\cal E}=\varepsilon\mp i\gamma$ with $\varepsilon, \gamma \in {\mathbb{R}}$ 
and the states with the $\mp$ sign, respectively, represent 
resonanc-decay and resonance-formation processes. 
This pairing property of the energy eigenvalues indicates that 
states in higher-dimensional PPB 
possibly have zero energy eigenstates. 
In fact the PPB in two dimensions (generally 
in even dimensions) has zero 
energy eigenvalue, which is included in the eigenvalues 
expressed by 
$\mp i(n_x-n_y)\hbar \gamma$. 
We see that the zero energy states are obtained for 
zero and positive integers satisfying $n_x=n_y$ 
and then they are infinitely degenerate. 
The zero energy states are interpreted as stationary 
flows round the centre of the PPB 
~\cite{sk4}. 
Furthermore, following hydrodynamics, it has been also shown that
some of such flows can be expressed by complex velocity 
potentials and various vortex structures appear 
 in the linear combinations of the 
 infinitely degenerate states. 
Considering that states in the conjugate spaces of GT are generally 
not normalizable but currents of those states are 
observable in quantum mechanics, 
the quantities observed in physical processes should be 
based on the probability currents such as currents in hydrodynamics. 
Hydrodynamical considerations will play a very important role 
in the investigations of quantum physics in GT. 
Hydrodynamical approach of quantum mechanics was vigorously 
investigated in the earlier stage of the development of 
quantum mechanics [9-16]. 
Vortices were extensively examined by 
Hirschfelder [17-20] 
and a review article was written by Ghosh and Deb~\cite{gd}. 
It should be noted that such hydrodynamical idea is 
still useful in present-day quantum physics~
\cite{sk4,bj,bb2}. 
Actually problems of vortices appear in many aspect of present-day 
physics such as vortex matters (vortex lattices)~\cite{blat,crab}, 
vortices 
in non-neutral plasma [26-29] 
and Bose-Einstein gases [30-34] 
 and so on. 
The vortex problems will hold  
a very important position in the hydrodynamical approach 
of quantum mechanics. 
As noted in ref.~\cite{sk4}, the stationary flows 
in the two-dimensional PPB can create almost arbitrary patterns of 
vortices because of the infinite degeneracy. 
PPB potentials can be a good approximation to the repulsive
 forces that are 
very weak at the centre of the forces such that harmonic oscillators 
are a good approximation to the attractive forces being very weak 
at the centre. 
In fact PPB has been 
applied in some chemical problems [35-37] 
The PPB, however, is a very special potential and then 
it seems to be difficult that the results of PPB extend to 
more general potentials.

In this article we shall investigate stationary flows 
in more general types of potentials from the hydrodynamical 
point of view. 
Especially, we study stationary flows in two dimensions 
discussed in the PPB~\cite{sk4}, 
because interesting quantities in hydrodynamics 
such as complex velocity potentials and vortices are definable 
in the dimensions.  
Furthermore it is well known that in two-dimensional 
hydrodynamical problems 
conformal mappings can be a very strong tool to investigate 
velocities, complex velocity potentials and vortices [38-41]. 
Some solutions solved in a special aspect are possibly 
extend to others in terms of conformal mappings. 
Particularly we will pay attention to 
the stationary flows that are described by
the zero energy eigenstates in the PPB~\cite{sk4}. 
Such zero energy states can also play an interesting role 
in statistical mechanics in GT, where a new type of entropy 
arises from the freedom with respect to 
the imaginary parts of energy eigenvalues [42-44]. 
That is to say, even if a many-particle system is 
in the ground state with a fixed energy 
and then it has no freedom arising from 
the real energy eigenvalue, it can still have 
the freedom with respect to the imaginary parts. 
Remembering the fact that all the energy eigenvalues in GT 
are expressed in the pairs of complex conjugates, 
we can understand the situation very easily, because 
the  imaginary part of every complex energy can be 
cancelled out by adding the conjugate energy 
in many-particle systems. 
An example was presented in ref.~\cite{ks6} for the PPB, 
where the burst of entropy from the new entropy was 
studied in thermal non-equilibrium. 
If we can find such zero energy states in more general 
types of potentials, we can investigate many-particle 
systems from a very new aspect where every state with 
a fixed energy can still have huge variety arising 
from the degeneracy of zero energy states. 
This article has two themes. 
One of the themes is to show the fact that rotational 
symmetric potentials of the type 
$V_a(\rho )=-a^2g_a \rho ^{2(a-1)/2}\ {\rm with}\ \rho=\sqrt{x^2+y^2}
\ {\rm and}\ a\not=0$  have the same zero energy 
solutions as those obtained in the PPB in two dimensions~\cite{sk4}. 
The other is vortex formation in terms of the zero energy solutions. 
We shall show that, as far as the zero energy solutions are concerned, 
Schr$\ddot {\rm o}$dinger equations 
with the rotational symmetric potentials $V_a(\rho )$ for all values of 
$a$ except$a=0$ 
are reduced to the same equation by using conformal mappings. 
We shall also see that the conformal mappings which are known to be 
very powerful tools in two-dimensional hydrodynamics 
become powerful tools also in the hydrodynamical approach 
of quantum mechanics and vortex patterns for all the potentials 
$V_a(\rho )$ can be investigated in a very simple method.

We shall perform our considerations as following; 
in section 2 general property of 
Schr$\ddot {\rm o}$dinger equations with rotational 
symmetric potentials is investigated in terms of 
conformal mappings. 
In section 3 it is shown that, for zero energy solutions, 
all the equations in the mapped spaces can 
be reduced to the same equation describing 
free motions in a constant potential. 
This means that, as far as the zero energy eigenstates are 
concerned, all the symmetric potentials have the same 
solutions with the infinite degeneracy 
as those obtained in the PPB~\cite{sk4}. 
Boundary conditions are also discussed there. 
Following the considerations of the PPB, hydrodynamical arguments 
are performed and velocity, complex velocity potentials and 
vortices are investigated in section 4. 
In section 5 an extension of the argument 
to three-dimensional problems are carried out and vortices 
 are studied in three dimensions. 
Non plane-wave solutions for the zero energy are briefly 
discussed in section 6. 
Remarks on non-zero energy solutions are presented 
in section 7. 
Some remarks and comments on the present work 
are given in section 8. 


\section{Conformal mappings of Schr\"{o}dinger equations with 
 symmetric  potentials} 
\label{sect.2.0}

We shall investigate the general structure of Schr\"{o}dinger equations 
$$
i\hbar\ {\partial  \over \partial t}\Psi(t,x,y)=H\Psi(t,x,y),
$$ 
where the Hamiltonian $H$ is described by 
rotational symmetric potentials in two-dimensional space 
$(x,y)$. 
The eigenvalue problems with the energy eigenvalue ${\cal E}$ 
are explicitly written as
\begin{equation}
 [-{\hbar^2 \over 2m}{\boldsymbol{\vartriangle}} +V_a(\rho)]\ \psi(x,y) 
  = {\cal E}\ \psi(x,y),
  \label{2.1.1}
\end{equation}
where $a \in {\mathbb{R}}$ ($a\not=0$), 
$$
{\boldsymbol{\vartriangle}}={\partial^2 \over \partial x^2}+{\partial^2 \over \partial y^2},
$$ 
$$
V_a(\rho)=-a^2 g_a\rho^{2(a-1)},
$$ 
with $\rho=\sqrt{x^2+y^2}$, 
$m$ and $g_a$ are, respectively, the mass of the particle and the coupling 
constant. 
Note here that $V_a$ represents repulsive potentials for $(g_a>0,\ a>1)$ and 
$(g_a<0,\ a<1)$ and attractive potentials for 
$(g_a>0,\ a<1)$ and $(g_a<0,\ a>1)$. 
Since we investigate the equations 
in the conjugate spaces of GT, 
the energy eigenvalues ${\cal E}$ of \eqref{2.1.1} are generally complex 
numbers.  

Following the hydrodynamical argument [38-41], 
let us consider the following conformal mappings; 
\begin{equation}
\zeta_a=z^a,\ \ \ \ \ \  {\rm with}\ z=x+iy.
\label{2.1.2}
\end{equation} 
Note that the conformal mappings are singular at the origin except 
in the cases for $a=$positive integers, 
and the conformal mapping for $a=1$ is trivial because nothing is 
changed by the mapping. 
We further notice that 
a complex factor $A$ can be multiplied in the mappings 
such as $\zeta=Az^a$, 
which will be discussed in 
the case for $A=e^{i\alpha}$ with $\alpha \in {\mathbb{R}}$. 
When we use the notation
$$
\zeta_a=u_a+iv_a,
$$ 
we see that  
\begin{equation} 
u_a=\rho^a\cos a\varphi,\ \ v_a=\rho^a\sin a\varphi,
\label{2.1.3}
\end{equation} 
where $u_a,\ v_a \in {\mathbb{R}}$ and 
$\rho=\sqrt{x^2+y^2},\ \varphi=\arctan (y/x)$. 
Using the notations 
$$
\rho_a^2=u_a^2+v_a^2(=\rho^{2a})\ \ {\rm and}\ \  \varphi_a=a\varphi, 
$$ 
we have  
\begin{equation}
u_a=\rho_a\cos \varphi_a\ \ {\rm and}\ \ v_a=\rho_a\sin \varphi_a.
\label{2.1.31}
\end{equation} 
In the $(u_a,v_a)$ plane the equations \eqref{2.1.1} are written down as 
\begin{equation}
a^2\rho_a^{2(a-1) / a} [-{\hbar^2 \over 2m}{\boldsymbol{\vartriangle}}_a-g_a]\ \psi(u_a,v_a)=
     {\cal E}\ \psi(u_a,v_a), 
 \label{2.1.4}
\end{equation}
where 
$$
{\boldsymbol{\vartriangle}}_a={\partial^2 \over \partial u_a^2}+{\partial^2 \over \partial v_a^2}.
$$ 
We can rewrite the equations as
\begin{equation}
[-{\hbar^2 \over 2m}{\boldsymbol{\vartriangle}}_a-g_a]\ \ \psi(u_a,v_a)=
    a^{-2}{\cal E}\ \rho_a^{2(1-a) / a} \psi(u_a,v_a). 
 \label{2.1.5}
\end{equation}   
Exchanging the second term of the left-hand side and the term of the
right-hand side, we obtain
\begin{equation}
[-{\hbar^2 \over 2m}{\boldsymbol{\vartriangle}}_a-
   a^{-2}{\cal E}\ \rho_a^{2(1-a) / a}]\ \psi(u_a,v_a)=
    g_a \ \psi(u_a,v_a). 
 \label{2.1.6}
\end{equation} 
It is quite interesting that we can read this equations as follows; 
the eigenvalue problem for the potential $V_a(\rho)$ in the $(x,y)$ plane 
given by \eqref{2.1.1} is replaced by the eigenvalue problem for 
the potential $V_{1/a}(\rho_a)$ in the $(u_a,v_a)$ plane, where 
the roles of the eigenvalue ${\cal E}$ and the coupling constant $g_a$ 
are exchanged. 
We may consider that this relation represents a kind of duality 
between the energy and the coupling constant. 
From the relation we see 
that by solving the eigenvalues for the fixed ${\cal E}$ 
in the $(u_a,v_a)$ plane we can determine the strength of the coupling 
constant $g_a$ to reproduce the eigenvalue ${\cal E}$ 
for the potential $V_a(\rho)$ in the $(x,y)$ plane. 
We shall return to the relations between the problems for 
$V_a$ in the $(x,y)$ plane and $V_{1/a}(\rho_a)$ in the $(u_a,v_a)$ plane 
in section 7, because this theme is not the main subject of this 
section. 

Here let us briefly comment on the conformal mappings $\zeta_a=z^a$. 
We see that the transformation maps the part of the $(x,y)$ plane 
described by $0\leq \rho<\infty, \ 0\leq \varphi<\pi/a$ 
on the upper half-plane of the $(u_a,v_a)$ plane for $a>0$ 
and the lower half-plane for $a<0$. 
Note here that the maps on the part of the $(u_a,v_a)$ plane with 
the angle $\varphi_a=\varphi-\alpha$ can be carried out 
by using the conformal mappings 
\begin{equation}
\zeta_a(\alpha)=z^a e^{-i\alpha}.
\label{2.1.7}
\end{equation}
In the maps the variables 
\begin{equation}
u_a(\alpha)=\rho^a \cos (a\varphi-\alpha)\ \ {\rm and}\ \  
v_a(\alpha)=\rho^a \sin (a\varphi-\alpha)
\label{2.1.8}
\end{equation} 
should be used. 
We also have the relations 
\begin{equation}
u_a(\alpha)=u_a\cos\alpha + v_a\sin\alpha \ \ {\rm and}\ \  
v_a(\alpha)=v_a\cos\alpha - u_a\sin\alpha. 
\label{2.1.9}
\end{equation} 
Of course, the relations $u_a(0)=u_a$ and $v_a(0)=v_a$ are obvious. 

It will be better to comment on the meaning of the choice of 
the variables $u_a$ and $v_a$ given in (3). 
It is obvious that $u_a$ and $v_a$ are not suitable variables 
to represent the states having definite properties with respect 
to rotations, such as the states with definite angular momentum, 
in comparison with the polar coordinates $\rho$ and $\varphi$. 
In the following discussions, however, we will be interested only 
in the states describing stationary flows that are basic elements 
in hydrodynamics. 
In general the stationary flows, such as those in scattering problems, 
cannot be described by the states with definite angular momentum, 
because every stationary flow 
has the specific directions representing the incoming and out-going flows. 
(Examples of the stationary flows in PPB will be presented in section 4. 
See figs. 1 and 2.) 
Such stationary flows, of course, have no definite rotational symmetry 
except rotations with respect to some specific angles. 
We can understand such situations by considering the fact that 
the directions of the 
incoming flows are chosen by hand in scattering experiments. 
Actually it will be shown that 
the freedom of the phase $\alpha$ in the conformal mapping 
(8) is related to such choices. 
(See section 3.) 
The choice of the variables $u_a$ and $v_a$ is, therefore, 
important in the following hydrodynamical approach, 
where the relations between the potentials 
$V_a(\rho )$ with different values of $a$ are studied. 
An explicit example of the difference between the choice of 
the polar coordinates and that of $u_a$ and $v_a$ has been shown 
in the case of PPB in section 3 of ref. [8].

\vskip10pt

\section{Zero energy  solutions of 
the Schr\"{o}dinger equations}
\label{sect.3.0}

We shall here study the special solutions having zero energy eigenvalue 
${\cal E}=0$.  
As noted in section 1, 
energy eigenvalues in GT are generally complex and 
all energy eigenvalues appear as pairs of  
complex conjugates like $\varepsilon \mp i\gamma$ ($\varepsilon,\gamma 
\in {\mathbb{R}}$)~\cite{bohm}. 
This indicates that, provided that a potential in one dimensional space has 
pure imaginary eigenvalues, the potentials extended in two dimensions 
possibly have zero energy states. 
This situation really occurs in parabolic potential barriers (PPB), 
that is, one dimensional PPB has pure imaginary eigenvalues [2-7] 
and hence two dimensional (generally in even dimensions) PPB 
has  zero energy states 
that are described by the stationary flows 
round the origin and infinitely degenerate~\cite{sk4}. 
Let us investigate zero energy solutions for the potentials 
$V_a(\rho)$.

\subsection{Zero energy solutions}
  \label{sect.3.3.0}

We see that for the zero energy ${\cal E}=0$ 
the Schr\"{o}dinger equations \eqref{2.1.5} 
obtained by the conformal mappings become very 
simple such that 
\begin{equation}
[-{\hbar^2 \over 2m}{\boldsymbol{\vartriangle}}_a-g_a]\ \psi(u_a,v_a)=0.  
 \label{3.1.1}
\end{equation}  
Note that the zero energy solutions have no time dependence. 
It is remarkable that the equation becomes same for all $a$, 
that is, the potential is expressed by the constant $g_a$ 
for all $a$. 
As far as the zero energy solutions are concerned, 
the equations transformed by the conformal mappings 
can be written in the same form for all the potentials 
$V_a(\rho )$ with $a\not=0$. 
It should  be noticed here that only 
in the case of the constant potential $V_1=-g_1$ for $a=1$ 
the energy eigenvalues can take arbitrary values satisfying 
th condition ${\cal E}+g_1>0$, because the right-hand side of (6) 
does not have any $\rho$ dependence. 

It is trivial that the solutions of (11) are given by
 the two-dimensional plane waves with the energy $g_a$. 
The solutions are, therefore, represented by 
\begin{equation}
\psi_0^\pm({\boldsymbol{\rho }}_a)=
N_a e^{\pm i{\boldsymbol{k}}_a(\theta)\cdot {\boldsymbol{\rho }}_a }, 
\ \ {\rm for}\ g_a>0 \label{3.1.2}
\end{equation} 
and 
\begin{equation}
\phi_0^\pm({\boldsymbol{\rho }}_a)=
M_a e^{\pm {\boldsymbol{k}}_a(\theta)\cdot {\boldsymbol{\rho }}_a}, 
\ \ {\rm for}\ g_a<0 \label{3.1.3}
\end{equation} 
where the angle $\theta$ denotes the moving direction of the plane wave 
in the $(u_a,v_a)$ plane, 
\hfil\break
${\boldsymbol{k}}_a(\theta)=\sqrt{2m|g_a|}/\hbar\cdot (\cos \theta,
\sin \theta)$ and 
 ${\boldsymbol{\rho }}_a=(u_a,v_a)$ are two-dimensional vectors, 
and $N_a$ and $M_a$ are 
in general complex numbers. 
Comparing the equation 
${\boldsymbol{k}}_a(\theta)\cdot {\boldsymbol{\rho }}_a
=k_a(u_a\cos \theta + v_a\sin \theta)$ where 
$k_a=\sqrt{2m|g_a|}/\hbar $ with $u_a(\alpha)$ of (10), 
we see that the angle $\theta$ can be adjusted to the phase $\alpha$ 
introduced in the conformal mappings (8). 
By using the phase $\alpha$ and the variable $u_a(\alpha )$  
the solutions (12) and (13) are written by 
\begin{equation}
\psi_0^\pm (u_a(\alpha ))=N_a e^{\pm ik_a u_a(\alpha ) }, 
\ \ {\rm for}\ g_a>0 \label{3.1.4}
\end{equation} 
and 
\begin{equation} 
\phi_0^\pm (u_a(\alpha ))=M_a e^{\pm k_a u_a(\alpha )}, 
\ \ {\rm for}\ g_a<0. \label{3.1.5}
\end{equation}  
We shall use the representations given in (14) and (15) 
in the following discussions. 
Note here that, taking acount of the relations 
\begin{equation}
u_a(\pm \pi/2)=\pm v_a,\ \ \ \ 
u_a(\pm \pi )=-u_a,  \label{3.1.6}
\end{equation} 
we can represent all the solutions of (14) and (15) by 
\begin{equation}
\psi_0^+(u_a(\alpha))\ \ \ {\rm and}\ \ \ \phi_0^+(u_a(\alpha))\ \ \ 
{\rm with} \ \ \ -\pi<\alpha\leq \pi.
\label{3.1.7}
\end{equation} 
We also notice that the solutions 
$\psi_0^\pm(u_a(\alpha))$ for $g_a>0$ are expressed by 
the plane waves with the fixed momentum $p_a=\sqrt{2mg_a}$, 
whereas those $\phi_0^\pm(u_a(\alpha))$ for $g_a<0$ are expressed by 
exponential growing or dumping functions. 
This difference is essential, because the plane-wave solutions can 
always be the states contained in the conjugate spaces of 
GT of which nuclear space is given by 
Schwartz space~\cite{bohm}, 
whereas the exponential growing functions such as 
$e^{\rho^a\cos (a\varphi-\alpha)}$ 
with $0<\cos (a\varphi-\alpha) $ cannot find a simple nuclear space 
for arbitrary values of $a$. 
From now on we shall mainly 
discuss on the plane-wave solutions for $g_a>0$. 

Let us summarize the main results in the cases of $g_a>0$. 
\hfil\break
(1) All the potentials written by $V_a(\rho)$ have zero energy 
eigenstates in GT. 
\hfil\break
(2) All the solutions with the zero energy can be expressed by 
the plane-wave with the fixed momentum $p_a=\sqrt{2mg_a}$  
in the $(u_a,v_a)$ plane. 
\hfil\break
(3) The zero energy solutions have an infinite freedom arising from 
the arbitrary angle $-\pi \leq\alpha<\pi$ that corresponds to 
the freedom of 
the angle between the incoming particle and the $x$ axis, 
which is given by $(\pi-\alpha)/a$. 
\hfil\break
(4) In the case 
of the constant potential corresponding to $a=1$, 
though we have the same solutions obtained in the above arguments,  
their energy eigenvalues need not equal zero 
but the energies can take arbitrary values 
fulfilling the relation ${\cal E}+g_1>0$.

\subsection{Property of the zero energy states for $2\pi$ rotation}
\label{sect.3.s.2}

In general eigenfunctions must have some 
definite properties with respect to the $2\pi$ rotation 
of the azimuthal angle $\phi$ 
 in the original $(x,y)$ plane corresponding to statistical properties 
 of the eigenstates, for instance, 
 $\psi(\theta,\phi+2\pi)=\psi(\theta,\phi)$ for integer-spin states 
 (bosonic states)
 and $\psi^\pm (\theta,\phi+2\pi)=-\psi^\pm (\theta,\phi)$ 
 for ${1 \over 2}$-spin states (fermionic states). 
 We see that in the latter condition (fermionic condition) 
 the eigenstates first return to the original states by $4\pi $ rotation. 
 It is known that states having new statistical properties which are 
different from Bose-Einstein and Fermi statistics appear  
in two dimensions and they are called anyon [45-49]. 
We may consider the eigenstates which first return to the original states 
by $2l\pi$ rotation ($l=$ integers $\geq3$), 
of which condition in two dimensions is expressed by 
$\psi^k(\phi+2\pi)=e^{i2k\pi/l}\psi^k(\phi)$ 
($k=$ positive integers $< l$) corresponding to anyon states. 
The condition may be called anyonic condition. 
Here we shall study the construction of the 
eigenfunctions under some different conditions. 
In this argument we represent the eigenfunctions 
in terms of 
$\psi_0^+(u_a(\alpha))\ \ {\rm and}\ \ \phi_0^+(u_a(\alpha))$ 
given in \eqref{3.1.4} and \eqref{3.1.5}. 

\hfil\break
(i) Cases for $a=\pm n \ \ {\rm with}\ n=1,2,3,\cdots$

Only the normal condition for the bosonic states (bosonic condition) 
\begin{equation}
\Psi_0^a(\rho, \varphi(\alpha)+2\pi)=\Psi_0^a(\rho, \varphi(\alpha))
 \label{4.1.1}
\end{equation} 
are available, 
because $\varphi(\alpha)=\varphi+\alpha/a$ and then  
$u_a(\alpha)$ do not change in 
the addition of $2a\pi$ to their phases for the choices 
\begin{equation}
a=\pm n \ \ {\rm with}\ n=1,2,3,\cdots.
\label{4.1.2}
\end{equation} 
All the eigenfunctions 
$\psi_0^+(u_a(\alpha))\ \ {\rm and}\ \ \phi_0^+(u_a(\alpha))$ 
including 
the freedom of the arbitrary angle $\alpha$ are satisfied by the 
bosonic condition. 

\hfil\break
(ii) Cases for $a=\pm(2n-1)/2 \ \ {\rm with}\ n=1,2,3,\cdots$

Two types of the conditions 
\begin{equation}
\Psi_0^a(\rho, \varphi(\alpha)+2\pi;\pm)
    =\pm \Psi_0^a(\rho, \varphi(\alpha);\pm)
\label{4.1.3}
\end{equation} 
are available. 
Since the addition of $2a\pi$ to the phase of $u_a(\alpha)$ 
changes only the signs of $u_a(\alpha)$  
for the choices 
$$
a=\pm(2n-1)/2 \ \ {\rm with}\ n=1,2,3,\cdots,
$$ 
we see that 
the linear combinations 
\begin{align}
\psi_0^{a}(u_a(\alpha);\pm)&=(\psi_0^\pm(u_a(\alpha))\pm 
        \psi_0^\mp(u_a(\alpha)))/\sqrt{2} \nonumber \\
\phi_0^{a}(u_a(\alpha);\pm)&=(\phi_0^\pm(u_a(\alpha))\pm 
        \phi_0^\mp(u_a(\alpha)))/\sqrt{2} 
        \label{4.1.4}
\end{align} 
fulfill the $\pm$ conditions, respectively.

\hfil\break
(iii) Cases for rational numbers

Let us here study the case 
for rational numbers written by irreducible fractional numbers 
$a_r=m/l$,  
which satisfies the relations $0<a_r< 1$. 
We can introduce generalized conditions 
 such that 
\begin{equation}
\Psi_0^{a_r}(\rho, \varphi(\alpha)+2\pi;e^{i 2k\pi/l})= 
      e^{i 2k\pi/l}\Psi_0^{a_r}(\rho, \varphi(\alpha);e^{i 2k\pi/l}),
\label{4.1.5}
\end{equation} 
where $k$ is zero or a positive integer being less than $l$. 
It is transparent that the cases for $(l=m=1)$ and $(l=2,m=1)$, 
respectively, 
correspond to the cases of (i) and (ii). 
The eigenfunctions satisfying the conditions are 
obtained as 
\begin{equation}
\psi_0^{a_r}(u_{a_r}(\alpha);\eta_k)=
  \sum_{n=0}^{l-1} (\eta_k)^{-n} \psi_0^+(u_a(\alpha+2na_r\pi)),
  \label{4.1.6}
\end{equation} 
where $0\leq  k<l$ and 
$$
\eta_k=e^{i2k\pi/l}. 
$$ 
Since the phase $\eta_k$ does not depend on the numerator $m$ of 
the rational number $a_r=m/l$, 
we see that the eigenfunctions for all $a_r$ 
with the same denominator $l$ can be expressed by the functions 
with the same property for the rotation of the angle $2\pi$ 
in the $(x,y)$ plane.  
We can, of course, describe 
the eigenfunctions for the case with $g_a<0$ 
by replacing $\psi_0^+(u_a(\alpha+2na_r\pi))$ 
with $\phi_0^+(u_a(\alpha+2na_r\pi))$ 
in the right-hand side of \eqref{4.1.6}.

Since all the rational numbers $q$ are expressed by 
$q= a_r+n$ with $n=0,\pm1,\pm2,\cdots$ 
and the addition of the integers to $a_r$ 
does not change the above argument at all, 
the eigenfunctions for $q$ satisfying the anyonic conditions are 
obtained by replacing $a_r$ with $q$. 
\vskip5pt

At this moment it is hard to answer the question 
whether the choice of irrational 
numbers for $a$ is physically meaningful or not. 
\vskip5pt

Note here that 
all the solutions described by the plane waves 
fulfill the same condition with respect to the $2\pi$ rotation 
in the $(u_a,v_a)$ plane, 
i.e., the bosonic condition, because  
the addition of $2\pi$ to the angle 
$\varphi_a$ does not change $u_a(\alpha)$. 
We see that in order to represent the whole $(u_a,v_a)$ plane 
the double sheets of the $(x,y)$ planes (Riemann surface) 
are needed for the choice of $a=\pm 1/2$. 
In general for the choice of $a_r=\pm 1/l$ 
the $l$ sheets of the $(x,y)$ plane 
like $l$ spiral sheets are required to 
cover the whole $(u_a,v_a)$ plane. 
We may consider that the case for $a=0$ can be understood 
in the limit of $l\rightarrow \infty$, where 
the infinite spiral sheets are needed in the $(x,y)$ plane.

\subsection{Infinite degeneracy of the zero energy states}
\label{sect.3.1.3}

A kind of degeneracy arising from the angle of the incoming 
particle with respect to the $x$ axis has been pointed out 
in section 2. 
We, however, see that the zero energy states have another 
type of infinite degeneracy that has been already solved 
in the two-dimensional PPB~\cite{sk4}. 
In the PPB the degeneracy arises from the pairing property of the 
energy eigenvalues given by $\mp i(n+1/2)\hbar \gamma$, that is, 
the energy eigenvalues of the type $\mp i(n_x-n_y)\hbar \gamma$ 
appear in the two-dimensional PPB and hence the infinitely degenerate 
zero energy states are derived for all the cases satisfying 
$n_x=n_y$. 
We see that the origin of the infinite degeneracy is due to 
the existence of the infinite number of resonances having 
the decay widths $(n+1/2)\hbar \gamma$ in the one-dimensional PPB and 
the coexistence of the resonance-formation and resonance-decay processes 
with equal probability in the two-dimensional PPB. 
The zero energy states are interpreted as the stationary flows
expressed by the incoming flows corresponding to the 
formation process and the out-going flows corresponding to 
the decay process, which will be shown in fig. 1 and fig. 2 of 
section 5. 
Let us see the degeneracy in (11) where the two-dimensional PPB 
is included. 
As an example we study the freedom for the wavefunction 
$\psi_0^\pm(u_a)$ given by~\eqref{3.1.2}. 
By putting the wavefunction $f^\pm (u_a;v_a)\psi_0^\pm(u_a)$ 
into~\eqref{3.1.1} 
where $f^\pm (u_a;v_a)$ is a polynomial function of $u_a$ and $v_a$, 
we obtain the equation 
\begin{equation}
[{\boldsymbol{\vartriangle}}_a \pm2ik_a{\partial \over \partial u_a}]f^\pm (u_a;v_a)=0.
\label{5.1.1}
\end{equation} 
As solved in ref.~\cite{sk4}, 
a few examples of the functions $f$ 
are given by  
\begin{align}
f_0^\pm (u_a;v_a)&=1, \nonumber \\
f_1^\pm (u_a;v_a)&=4k_a v_a, \nonumber \\
f_2^\pm (u_a;v_a)&=4(4k_a^2 v_a^2+1\pm 4 i k_a u_a).
\label{5.1.2}
\end{align} 
In the two-dimensional PPB 
the functions are generally written by the multiple of 
the polynomials of degree $n$, $H_n^\pm(\sqrt{2k_2} x)$, 
such that 
\begin{equation}
f_{n}^{\pm}(u_2;v_2)=H_{n}^\pm(\sqrt{2k_2} x)
\cdot H_{n}^\mp(\sqrt{2k_2} y),
\label{5.1.3} 
\end{equation} 
where $x$ and $y$ in the right-hand side should be 
considered as the functions of $u_2$ and $v_2$~\cite{sk4}. 
Since the form of the equations~\eqref{5.1.1}  
is the same for all $a$, 
the solutions can be written by the same polynomial functions 
that are given in~\eqref{5.1.3} for the PPB. 
That is to say, 
we can obtain the polynomials for arbitrary $a$ 
by replacing $u_2$ and $v_2$ with $u_a$ and $v_a$ 
in~\eqref{5.1.3}. 
Note that the polynomials $H_{n}^\pm(\xi)$ with 
$\xi=\sqrt{m\gamma/\hbar} x$ 
are defined 
by the solutions for the eigenstates with 
${\cal E}=\mp i(n+1/2)\hbar \gamma$ in one dimensional 
PPB of the type $V(x)=-m\gamma^2x^2/2$ and they are 
written in terms of the Hermite polynomials $H_{n}(\xi)$ as 
\begin{equation}
H_{n}^\pm(\xi)=e^{\pm i n\pi/4} H_{n}(e^{\mp i\pi/4}\xi). 
\label{5.1.4}
\end{equation} 
(For details, see ref.~\cite{cdlp,sk}.) 
It is remarkable that all the wavefunctions for arbitrary $a$ 
can be represented by the same functions of the PPB 
in the $(u_a,v_a)$ plane. 
For $\psi_0^\pm(v_a)$ we should take the polynomials 
$f_n^\pm (v_a;u_a)$ in which the variables $u_a$ and $v_a$ are 
exchanged. 
 
Let us here briefly comment on the boundary conditions 
discussed in the last subsection. 
Considering the relations 
$u_a(\alpha+2\pi)=u_a(\alpha)$ and $v_a(\alpha+2\pi)=v_a(\alpha)$, 
we can make the eigenstates satisfying the suitable 
boundary conditions by using the eigenfunctions 
$\psi_{0n}^+(u_a(\alpha))=f_n^+ (u_a(\alpha );v_a(\alpha ))
\psi_0^+(u_a(\alpha ))$ 
in stead of $\psi_0^+(u_a(\alpha ))$ in (3.7), \eqref{4.1.4} and 
\eqref{4.1.6}. 
Note also that the eigenfunctions 
$\psi_{0n}^+(u_a(\alpha))$ 
for $n\geq 1$ does not describe 
plane waves and hence they cannot be normalized in terms of 
$\delta$ functions. 
We essentially have to treat them as the eigenfunctions of 
the conjugate space ${\cal S}({\mathbb{R}}^2)^\times $ in GT that is expressed by 
$$
{\cal S}({\mathbb{R}}^2)\subset {\cal L}^2({\mathbb{R}}^2) \subset {\cal S}({\mathbb{R}}^2)^\times ,
$$ 
where ${\cal S}({\mathbb{R}}^2)$ and ${\cal L}^2({\mathbb{R}}^2)$ 
are, respectively, Schwartz space and Lebesgue space 
in two dimensions. 
(For details, see~\cite{bohm,cdlp,sk,sk4}.) 
We will see in the next section 
that this degeneracy plays an important role to investigate vortices. 

Note that we can obtain the 
polynomials for the wavefunctions $\phi_0^\pm(u_a)$ 
by replacing $k_a$ with $-i k_a$ in the polynomials 
derived from~\eqref{5.1.3}. 
We also easily see that in one dimension the equation corresponding 
to~\eqref{5.1.1} does not bring any new freedom 
to the plane-wave solutions.

\section{Hydrodynamical considerations of the zero energy states}
\label{sect.4.0}

In hydrodynamics conformal mappings are very powerful tools 
to understand structures of currents. 
Actually the important hydrodynamical ideas such as 
the property of complex velocity potentials, circulations of currents, 
strengths of vortices, 
strengths of sources and so forth do not change in the conformal 
mappings [38-41]. 
This fact means that we can simultaneously carry out 
the investigation of the hydrodynamical 
properties of the zero energy solutions for all the potentials 
$V_a(\rho)$ in the mapped spaces, i.e., in the $(u_a,v_a)$ plane. 
Results for all the potentials with $a\not=0$ 
can be obtained by the inverse 
transformations of the conformal mappings. 
In this section we shall study the zero energy states  
from a hydrodynamical viewpoint for the  $g_a>0$ cases, 
because the eigenstates for $g_a<0$ represented by exponential 
growing or dumping functions do not describe any oscillating 
waves, which will be briefly discussed in section 6. 

\subsection{Currents and velocities}
\label{sect.4.1.1}

Though states in GT are generally not normalizable, 
the probability currents are observable  in physical processes 
such as in scattering processes. 
We shall, therefore, study the currents and other quantities 
based on hydrodynamics. 
The probability density $\rho(t,x,y)$ and 
the probability current ${\boldsymbol{j}}(t,x,y)$ of a state $\psi(t,x,y)$ 
in non-relativistic quantum mechanics 
are defined  
by 
\begin{align}
  \rho(t,x,y)&\equiv\left| \psi(t,x,y)\right|^2, \label{6.1.1}\\
  {\boldsymbol{j}}(t,x,y)&\equiv{\rm Re}\left[\psi(t,x,y)^*
  \left(-i\hslash\nabla\right)\psi(t,x,y)\right]/m.
  \label{6.1.2}
\end{align}
They satisfy the equation of continuity 
\begin{equation}
  \frac{\partial\rho}{\partial t}+\nabla\cdot{\boldsymbol{j}}=0. 
   \label{6.1.3}
\end{equation}
Following the analogue of the hydrodynamical 
approach [38-41], 
the fluid 
can be represented by 
the density $\rho$ and the fluid velocity ${\boldsymbol{v}}$. 
They satisfy Euler's equation of continuity 
\begin{equation}
  \frac{\partial\rho}{\partial t}+\nabla\cdot(\rho{\boldsymbol{v}})=0. 
   \label{6.1.4}
\end{equation} 
Comparing this equation with \eqref{6.1.3}, 
we are thus led to the following 
definition for the quantum velocity of a state $\psi(t,x,y)$, 
\begin{equation}
  {\boldsymbol{v}}\equiv\frac{{\boldsymbol{j}}(t,x,y)}{\left| \psi(t,x,y)\right|^2}, 
   \label{6.1.5}
\end{equation}
in which ${\boldsymbol{j}}(t,x,y)$ is given by \eqref{6.1.2}. 
Notice that $\rho$ and ${\boldsymbol{j}}$  for the zero energy states 
do not depend on time $t$. 

Let us discuss them in the $(u_a,v_a)$ plane. 
All quantities $O$ defined in the $(u_a,v_a)$ plane 
will be marked by the symbol `hat' such as $\hat O$ that can easily 
be transformed into the quantities in the $(x,y)$ plane. 
It is apparent that in the $(u_a,v_a)$ plane the currents 
of the plane waves $\psi_0^+(u_a(\alpha))$ 
are represented by the same form for all $a$ 
\begin{align}
\hat{j}_{u_a}&=|N_a|^2\hbar k_a\cos\alpha/m,  \ \  \nonumber \\
\hat{j}_{v_a}&=|N_a|^2\hbar k_a\sin\alpha/m.
\label{6.1.6} 
\end{align} 
Note here the following relations; 
$$u_a(\alpha)=u_a \cos\alpha +v_a \sin\alpha,\ \ 
u_a(0)=u_a,\ \ v_a(0)=v_a.$$ 
When we represent the momentum in terms of 
the vector of the $(u_a,v_a)$ plane as 
$$
\hat{{\boldsymbol{p}}}_{a}=(\hbar k_a\cos\alpha,\hbar k_a\sin\alpha)
$$ 
for $ \psi_0^+(u_a(\alpha))$, 
the currents are generally written by 
\begin{equation}
\hat{{\boldsymbol{j}}}=|N_a|^2\hat{{\boldsymbol{p}}}_{a}/m. 
\label{6.1.7}
\end{equation}
Hence the velocities are given by 
\begin{equation}
\hat{{\boldsymbol{v}}}=\hat{{\boldsymbol{p}}}_{a}/m. 
\label{6.1.8}
\end{equation} 
Following the argument of hydrodynamics (for details, see Appendix 
of ref.~\cite{sk4}), we can introduce the complex velocity potential 
$W$ as 
\begin{equation}
W_a=(\hat{p}_{u_a}-i\hat{p}_{v_a})\zeta_a/m. 
\label{6.1.9}
\end{equation} 
The velocity potential $ \varPhi$ and the stream 
function $ \varPsi$ can be introduced as same as those 
in hydrodynamics by 
$$
W_a=\varPhi+i\varPsi,
\label{6.1.10}
$$ 
where they satisfy the following relations in the $(u_a,v_a)$ plane; 
\begin{equation}
   \hat v_{u_a} =\frac{\partial {\boldsymbol{\varPhi}}}{\partial u_a}
   =\frac{\partial {\boldsymbol{\varPsi}}}{\partial v_a}, \ \ \ \ 
   \hat v_{v_a} =\frac{\partial {\boldsymbol{\varPhi}}}{\partial v_a}
   =-\frac{\partial {\boldsymbol{\varPsi}}}{\partial u_a}. 
   \label{6.1.11}
\end{equation}
It is known that Cauchy-Riemann's equations are satisfied 
by the velocity potential and the stream function.

The velocities of the 
$u_a$ and $v_a$ directions in the $(x,y)$ plane are given by 
\begin{equation}
   v_{u_a} =h_a \hat v_{u_a} 
    \ \ \ \ 
   v_{v_a} =h_a \hat v_{v_a},
   \label{6.1.13}
\end{equation} 
where $h_a=a(u_a^2+v_a^2)^{(a-1)/2a}$. 
Hydrodynamics tells us that $W_a$ describes corner flows 
with the angle $\pi/a$ round the origin. 
For example, in the case of the PPB with $a=2$~\cite{sk4} 
the plane waves in the $(u_a,v_a)$ plane, $\psi_0^\pm(u_2)$, 
are expressed in fig.~\ref{fig:4.1} and fig.~\ref{fig:4.2}. 
 \begin{figure}[pb]
   \begin{center}
    \begin{picture}(200,200)
     \thicklines
     \put(0,100){\vector(1,0){200}}
     \put(100,0){\vector(0,1){200}}
     \put(90,88){$0$}
     \put(205,98){$x$}
     \put(98,205){$y$}
     
     \qbezier(105,200)(105,105)(200,105)
     \qbezier(95,200)(95,105)(0,105)
     \qbezier(95,0)(95,95)(0,95)
     \qbezier(105,0)(105,95)(200,95)
     
     \put(200,105){\vector(1,0){1}}
     \put(0,105){\vector(-1,0){1}}
     \put(0,95){\vector(-1,0){1}}
     \put(200,95){\vector(1,0){1}}
    \end{picture}
   \end{center}
   \caption[]{Corner flows for $\psi_0^+(u_a)$ in two-dimensional PPB.}
   \label{fig:4.1}
  \end{figure}
  \begin{figure}
   \begin{center}
    \begin{picture}(200,200)
     \thicklines
     \put(0,100){\vector(1,0){200}}
     \put(100,0){\vector(0,1){200}}
     \put(90,88){$0$}
     \put(205,98){$x$}
     \put(98,205){$y$}
     
     \qbezier(105,200)(105,105)(200,105)
     \qbezier(95,200)(95,105)(0,105)
     \qbezier(95,0)(95,95)(0,95)
     \qbezier(105,0)(105,95)(200,95)
     
     \put(105.5,200){\vector(0,1){1}}
     \put(95.5,200){\vector(0,1){1}}
     \put(95.5,0){\vector(0,-1){1}}
     \put(105.5,0){\vector(0,-1){1}}
    \end{picture}
   \end{center}
   \caption[]{Corner flows for $\psi_0^-(u_a)$ in two-dimensional PPB.}
   \label{fig:4.2}
  \end{figure} 
Note that the states multuplied by the polynomials 
$f_0^\pm $ and $f_1^\pm $ of~\eqref{5.1.2} 
also represent the corner flows 
with the angle $\pi /a$.

\subsection{Vortices in the zero energy states}
\label{sect.4.1.2}

In hydrodynamics vortices are very important objects. 
In quantum mechanics, 
since the velocity defined by \eqref{6.1.5} diverges 
at the zero points of the wavefunctions, 
the vortices generally 
appear at such nodal points of 
the wavefunctions [17-20]. 
The situation is, however, not so simple to determine the positions 
of vortices, because the vortices do not always 
appear at the points where 
the wavefunctions vanish, when the currents also vanish at the 
same points. 
Since the zero energy states have an infinite degeneracy and also 
the freedom of the angle $\alpha $, 
we will be able to create vortex patterns having arbitrary 
number of vortices at arbitrary positions. 
General study of quantized vortices is carried out in ref.~\cite{joh5}. 
We shall here discuss the vortex patterns in a few simple cases 
of the linear combinations in terms of the infinite degeneracy. 

Let us study the vortex structures appearing in the 
linear combinations of two eigenstates constructed from 
\eqref{3.1.4} and \eqref{5.1.2}. 
The following discussions are carried out in the $(u_a,v_a)$ 
plane, because the singularities of the velocity 
does not change in the conformal mappings 
 except the singularity of the mappings at origin 
 for $a\not=$positive integers. 
The general form of the linear combination of two states 
can be written as 
\begin{equation}
\Psi=\psi_1+\psi_2, 
\label{5.2.1}
\end{equation} 
where, since the two states are not normalized, 
 the complex coefficients appearing 
in the linear combination are included in the two wavefunctions
$\psi_1$ and $\psi_2$. 
The absolute square of $\Psi$ is evaluated as 
\begin{equation}
|\Psi|^2=|\psi_1|^2+|\psi_2|^2+2{\rm Re}(\psi_1^*\psi_2). 
\label{5.2.2}
\end{equation} 
In general a component of the current of $\Psi$ is written as 
\begin{equation}
\hat{j}_\mu  ={\hbar \over m}{\rm Re}[\Psi^* (A_\mu \psi_1+B_\mu \psi_2)],
\label{5.2.3}
\end{equation} 
where $\mu=u_a$ or $v_a$, and $A_\mu $ and $B_\mu $ are complex functions 
defined by 
\begin{equation}
A_\mu=-i{\partial \psi_1 \over \partial \mu} \psi_1^{-1}\ \ \ 
B_\mu=-i{\partial \psi_2 \over \partial \mu} \psi_2^{-1}.
\label{5.2.10}
\end{equation} 
Let us study the nodal points of $|\Psi|^2$, where the vortices appear. 
We have 
\begin{equation}
|\Psi|^2=|\psi_1|^2+|\psi_2|^2+2|\psi_1||\psi_2|\cos \theta,  
\label{5.2.4}
\end{equation} 
where $\theta$ denotes the phase between $\psi_1$ and $\psi_2$. 
It is trivial that 
nodal points appear when the following two conditions are 
fulfilled;
\begin{equation}
|\psi_1|=|\psi_2|\ \ \ {\rm and}\ \ \ \cos\theta=-1. 
\label{5.2.5}
\end{equation} 
We put the first relation into 
\eqref{5.2.4} and thus  
obtain 
\begin{equation}
|\Psi|^2=2|\psi_1|^2(1+\cos \theta).  
\label{5.2.6}
\end{equation}
Taking account of the same relation $|\psi_1|=|\psi_2|$, 
the current is written by
\begin{align}
\hat{j}_\mu  &={\hbar \over m}|\psi_1|^2
    [(|A_\mu|\cos\phi_A+|B_\mu|\cos\phi_B)(1+\cos\theta)  \nonumber \\ 
    &+(|A_\mu|\sin\phi_A-|B_\mu|\sin\phi_B)\sin\theta],
\label{5.2.7}
\end{align} 
where $\phi_A$ and $\phi_B$ are, respectively, the phases of $A_\mu$ 
and $B_\mu$. 
The velocity is evaluated as 
\begin{align}
\hat{v}_\mu&={\hbar \over 2m}
    [(|A_\mu|\cos\phi_A+|B_\mu|\cos\phi_B)   \nonumber \\ 
    &+(|A_\mu|\sin\phi_A-|B_\mu|\sin\phi_B){\sin\theta \over 1+\cos\theta}].
\label{5.2.8}
\end{align} 
We see that the second term in the bracket [......] diverges 
by de l'Hospital's theorem, 
when the second condition for the angle,  $\cos\theta=-1$, 
is fulfilled. 
Thus we can obtain the condition for the divergence of the 
velocity 
\begin{equation}
|A_\mu|\sin\phi_A-|B_\mu|\sin\phi_B\not=0.
\label{5.2.9}
\end{equation} 
This equation means that the functions $A_\mu $ and $B_\mu $ 
must not be real  and 
also the imaginary parts of $A_\mu$ and $B_\mu$ must not be equal 
at least for one of the components $\mu=u_a \ {\rm and}\ v_a$. 

Now we can summarize the conditions for 
the determination of the vortex positions 
in the linear combinations of two wavefunctions 
$\psi_1$ and $\psi_2$ as follows:
\hfil\break
(I) $|\psi_1|=|\psi_2|$,
\hfil\break
(II) $\theta=(2l-1)\pi$ with $l$=integers,
($\theta$ is the phase between 
$\psi_1$ and $\psi_2$),
\hfil\break
(III) $|A_\mu |\sin\phi_A-|B_\mu |\sin\phi_B\not=0$, 
($A_\mu $ and $B_\mu $ are defined 
in \eqref{5.2.10}).
\hfil\break
Let us investigate the above conditions in a few simple examples.

\hfil\break 
{\bf Example (i)}: It is trivial that any linear combinations composed of 
the wavefunctions with the lowest polynomial (25) 
have no vortex, because the condition (III) is not fulfilled 
whereas nodal points satisfying the conditions (I) and (II) appear 
in the linear combinations. 

\hfil\break
{\bf Example (ii)}: The combination of the lowest polynomial and 
the second one 
such that 
$$\Psi=\psi_0^+(u_a(\alpha))-
          Cf_1^+(u_a(0);v_a(0))\psi_0^+(u_a(0))$$ 
has vortices at positions fulfilling the following conditions 
derived from (I) and (II);
\begin{align}
v_a(0)&=(-1)^n /4|C|k_a,  \nonumber \\ 
\hat{\theta}+\theta_C&=n\pi, \ \ \ \ \ \ \ \ \ \ (n={\rm integers}),
\label{5.2.11}
\end{align}
where $\theta_C$ is the phase of $C$ and 
\begin{align}
\hat{\theta}&=k_a[u_a(0)-u_a(\alpha)] \nonumber \\
      &=k_a[u_a(0)(1-\cos \alpha)-v_a(0)\sin \alpha].
\label{5.2.12}
\end{align} 
Let us examine the relations \eqref{5.2.11} in two cases for 
$a=1$ and $2$, where $C$ is taken to be a real number, i.e., 
$\theta_C=0$.

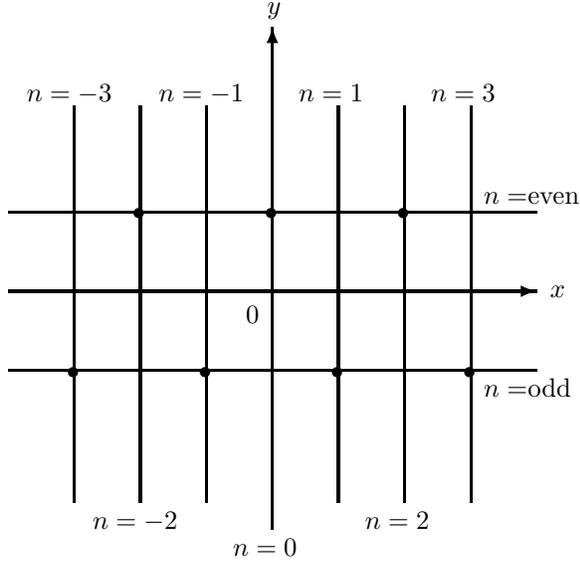
\begin{figure}
   \begin{center}
    \begin{picture}(200,200)
     \thicklines
     \put(0,100){\vector(1,0){200}}
     \put(100,10){\vector(0,1){190}}
     \put(90,88){$0$}
     \put(205,98){$x$}
     \put(98,205){$y$}
     \put(85,0){$n=0$}
     
     \put(0,70){\line(1,0){200}}
     \put(0,130){\line(1,0){200}}
     
     \put(25,20){\line(0,1){150}}
     \put(50,20){\line(0,1){150}}
     \put(75,20){\line(0,1){150}}
     \put(125,20){\line(0,1){150}}
     \put(150,20){\line(0,1){150}}
     \put(175,20){\line(0,1){150}}
     
     \put(180,133){$n=$even}
     \put(180,60){$n=$odd}
     
     \put(7,172){$n=-3$}
     \put(32,10){$n=-2$}
     \put(57,172){$n=-1$}
     \put(110,172){$n=1$}
     \put(135,10){$n=2$}
     \put(160,172){$n=3$}
     
     \put(47,127){$\bullet$}
     \put(97,127){$\bullet$}
     \put(147,127){$\bullet$}
     \put(22,67){$\bullet$}
     \put(72,67){$\bullet$}
     \put(122,67){$\bullet$}
     \put(172,67){$\bullet$}
    \end{picture}
   \end{center}
   \caption[]{Positions of vortices for $n=0,\ \pm 1,\ \pm 2,\ \pm 3$ 
   in a constant potential ($a=1$), 
   which are denoted by $\bullet$.} 
   \label{fig:4.3}
  \end{figure}

\hfil\break
Case for $a=1$: 
In this case we have $u_1(0)=x$ and $v_1(0)=y$ 
and then the relations are reduced to 
\begin{align}
y=(-1)^n /4|C|k_1& , \nonumber \\
x(1-\cos\alpha)-y\sin\alpha&=n\pi/k_1.
\label{5.2.13}
\end{align}
All vortices appear  on the two lines $y=\pm 1/4|C|k_1$ 
parallel to the $x$ axis 
and they are at the cross points of the two lines and the lines 
$x=(n\pi+(-1)^n \sin\alpha/4|C|)/k_1 (1-\cos \alpha)$ 
for $\alpha\not=0$. 
The positions of vortices for $n=0,\pm 1,\pm 2,\pm 3$ are presented 
in fig.~\ref{fig:4.3}, where $\alpha=\pi$ is taken. 
This situation is quite similar to the vortices called parallel 
vortex lines obtained in hydrodynamics. 
  
\begin{figure}
   \begin{center}
    \begin{picture}(200,200)
     \thicklines
     \put(0,100){\vector(1,0){200}}
     \put(100,0){\vector(0,1){200}}
     \put(93,85){$0$}
     \put(205,98){$x$}
     \put(98,205){$y$}
     
     \put(10,10){\line(1,1){180}}
     \put(10,190){\line(1,-1){180}}
     
     \qbezier(14,10)(100,96)(186,10)
     \qbezier(14,190)(100,104)(186,190)
     
     \qbezier(10,14)(96,100)(10,186)
     \qbezier(190,14)(104,100)(190,186)
     
     \qbezier(105,200)(105,105)(200,105)
     \qbezier(95,200)(95,105)(0,105)
     \qbezier(95,0)(95,95)(0,95)
     \qbezier(105,0)(105,95)(200,95)
     
     \put(125.8,125.8){$\bullet$}
     \put(68.5,68.5){$\bullet$}
     \put(81.8,146){$\odot$}
     \put(111,49.0){$\odot$}
     \put(145.5,82.0){$\diamond$}
     \put(49.5,112.6){$\diamond$}
     \end{picture}
   \end{center}
   \caption[]{Positions of vortices for $n=0,\ \pm 1$ in PPB ($a=2$), 
   which are denoted by $\bullet$ for $n=0$, $\diamond$ for $n=1$ and 
   $\odot$ for $n=-1$, respectively.}
   \label{fig:4.4}
  \end{figure}
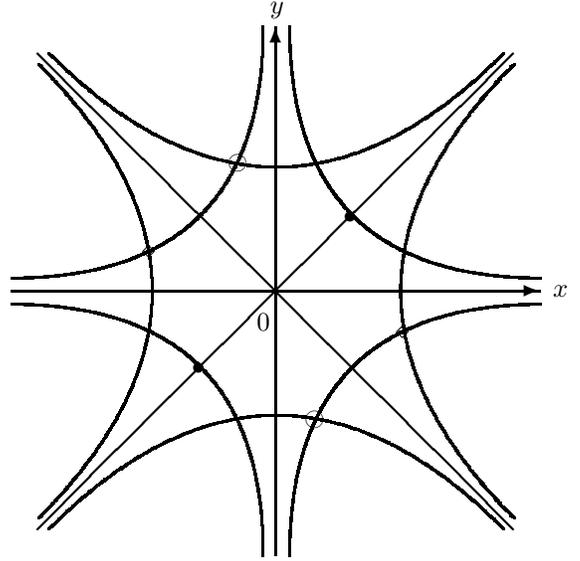

\hfil\break
Case for $a=2$ (PPB): 
Since the inverse transformation of the conformal mapping 
is described by the equations $u_2(0)=x^2-y^2$ and $v_2(0)=2xy$ 
in PPB~\cite{sk4}, the relations are given by 
\begin{align}
2xy&=(-1)^n /4|C|k_2, \nonumber \\
(x^2&-y^2)(1-\cos\alpha)-2xy\sin\alpha=n\pi/ k_2.  
\label{5.2.16}
\end{align} 
Vortices appear at the cross points of $x^2-y^2=(n\pi+
  (-1)^n \sin \alpha/4|C|)/k_2 (1-\cos \alpha)$ and 
  $xy=(-1)^n /8|C|k_2$. 
The positions of two vortices for $n=0$ and other four for 
$n=\pm 1$ are figured in fig.~\ref{fig:4.4}, 
where $\alpha=\pi$ is taken. 
The vortices appear at the symmetric positions with respect to 
the origin, which are described by the cross points of 
the two equations; 
\begin{equation}
x^2-y^2=n\pi/2k_2, \ \ \ \ 
xy=(-1)^n /8|C|k_2. 
\label{5.2.17}
\end{equation} 
We can make so many variety of the vortex patterns by changing of 
the parameters, $\alpha$ and $C$, and the zero energy states  
in terms of the polynomials \eqref{5.1.2}. 
Here we stress that, as shown in the above discussions, 
the higher polynomial solutions with $n\not=0$ 
which are not described by the plane waves in the $(u_a,v_a)$ plane 
play the essential roles to create vortices. 
\vskip5pt

Note here that the strength of vortex is characterized 
by the circulation $\varGamma$ 
that is represented by the integral round a closed contour $C$
encircling the vortex such that 
\begin{equation}
\varGamma=\oint_C {\boldsymbol{v}} \cdot d{\boldsymbol{s}}
\label{5.2.18}
\end{equation} 
and it is quantized as 
\begin{equation}
\varGamma=2\pi l\hbar/m, 
\label{5.2.19}
\end{equation} 
where the circulation number $l$ is 
an integer~\cite{joh2,joh5,bb2}. 
After simple but tedious calculations, we obtain that $l=-1$ for  
the vortices with $n=$ even and $l=1$ for the vortices with $n=$ odd. 

Before closing this section we point out the fact that we can 
realize almost all of vortex patterns because of the infinite 
degeneracy of the zero energy solutions. 
The study of the vortex patterns will be carried out 
by determining the parameter $a$ (the type of potential) 
and by finding the best linear combination in terms of the 
infinitely degenerate zero-energy-states to describe the 
vortex patterns. 
\vskip10pt

\section{Vortices in three dimensions}
\label{sect.4.3}

We shall study vortices in three dimensions. 
It is obvious that the conformal mappings given in \eqref{2.1.2} 
cannot apply in three dimensions. 
Schr\"{o}dinger equations in three dimensions are 
generally written as 
\begin{equation}
 [-{\hbar^2 \over 2m}({\boldsymbol{\vartriangle}} +{\partial^2 \over \partial z^2})
 +V_a(x,y,z)]\ \psi(x,y,z) 
  = {\cal E}\ \psi(x,y,z), 
  \label{4.3.1}
\end{equation} 
where ${\boldsymbol{\vartriangle}}=\partial^2/ \partial x^2+\partial^2/ \partial y^2$. 
The equations, however, can be reduced to two-dimensional ones 
in cases where potentials are separable into $2+1$ dimensions 
and the zero energy solutions are 
applicable to such cases. 
We shall discuss in the following three cases. 

\hfil\break
(I) Cases of free motions
 \hfil\break
Let us consider the cases where the motion of one direction 
(say $z$) is a free motion described by 
$e^{ik_z z}$. 
In these cases the potentials can be written by two dimensional 
potentials such as $V_a(\rho )$. 
In such cases we have 
\begin{equation}
 [-{\hbar^2 \over 2m}{\boldsymbol{\vartriangle}} +V_a(\rho)]\ \psi(x,y)e^{ik_z z} 
  = ({\cal E}-E_z)\ \psi(x,y)e^{ik_z z},
  \label{4.3.2}
\end{equation}
with $E_z=\hbar^2 k_z^2/2m$. 
The zero energy states in the two dimensions of $(x,y)$ 
appear when the relation 
\begin{equation}
{\cal E}-E_z=0
\label{4.3.3}
\end{equation} 
is fulfilled. 
It is trivial that the total energy ${\cal E}$ must be 
positive because of $E_z>0$. 
We see that all the vortex patterns  discussed in 
the last section appear 
in three-dimensional phenomena where 
the motion of one direction perpendicular to the vortex plane 
(the $(x,y)$ plane) is a free motion. 
Vortices in a constant magnetic field of $z$ direction 
will be one of these cases. 

\hfil\break
(II) Cases of exponentials
 \hfil\break
We can present another cases where the motions perpendicular 
to the vortex plane are described by exponentials such that 
$\psi(z) =e^{k_z z}$. 
The equation \eqref{4.3.1} is written as 
\begin{equation}
 [-{\hbar^2 \over 2m}{\boldsymbol{\vartriangle}} +V_a(\rho)]\ \psi(x,y)e^{k_z z} 
  = ({\cal E}+E_z)\ \psi(x,y)e^{k_z z}. 
  \label{4.3.4}
\end{equation} 
When the relation 
\begin{equation}
{\cal E}+E_z=0
\label{4.3.5}
\end{equation} 
is fulfilled, 
we have the same zero energy solutions 
and then we obtain the same vortex structures of the two dimensions. 
In these cases the total energy ${\cal E}$ must apparently be zero or 
negative.  

\hfil\break
(III) Cases for separable potentials
 \hfil\break
Even if the potential of the $z$ component is not constant potentials, 
the eigenvalue equations are separable into the $(x,y)$ and the $z$-
components 
in the cases with potentials such that 
\begin{equation} 
V(x,y,z)=V_2(x,y)+V_1(z).
\label{4.3.6}
\end{equation}  
When the eigenvalues of the $z$ component are given by $E_z$, 
the equations for the $(x,y)$ component become same as those 
of the cases (I) and (II).

In these three cases where the vortex plane $(x,y)$ and the other axis $(z)$ 
perpendicular 
to the vortex plane are completely separable 
and then the wavefunctions are written by the multiplicative forms such as  
$\psi(x,y)\psi(z)$, 
all vortices are described by the axial type and 
the toroidal  vortices do not appear~\cite{joh5}, 
because the positions of the vortices in the $(x,y)$ plane do not depend on $z$. 
\vskip5pt

Here we would like to note the construction of vortices 
by plane-wave solutions in the $(x,y)$ plane. 
Let us put the functions 
\begin{align}
\psi_0(x,y,z)&=N_a e^{ i(k_x x +k_y y)}(e^{i k_z z} \ {\rm or}\ 
e^{k_z z})\ \ \nonumber \\
&\ {\rm for} \ \ k_x,k_y,k_z \in {\mathbb{R}} 
\label{5.2.14}
\end{align}
 into \eqref{4.3.1}, in which $V_a(x,y,z)=0$ is taken. 
In this case, however, the relations for zero energy \eqref{4.3.3} 
and \eqref{4.3.5} 
should not be taken but 
the different relations ${\cal E}\mp E_z>0$ must be required. 
Taking $\hbar^2 k^2/2m={\cal E}\mp E_z$, 
we have the solutions same as those for the constant potential 
$V_2=-g_a$ in \eqref{4.3.6}, 
where $g_a=\hbar^2 k^2/2m$ and then $k_x^2+k_y^2=k^2$. 
This fact implies that the parallel vortices discussed in section 
4.2 are producible from the plane wave and the degenerate states 
with non-zero energy in three dimensions. 

Real vortex phenomena [24-31] appear 
in three dimensional spaces. 
Some of the vortex phenomena will be understood in the 
cases discussed above.

\vskip10pt

\section{Short notes on zero energy solutions for $g_a<0$}
\label{sect.5.0}

As noted in the section 3.1, we have the zero 
energy solutions $\phi_0^\pm(u_a(\alpha ))=M_a e^{\pm k_au_a(\alpha )}$ 
of (15). 
In general they are unnormalizable in the $(x,y)$ plane. 
In some special cases, however, some of the four can be 
normalizable. 
For example, provided that the parameters $a$ and $\alpha$ are taken 
so as to fulfill the relation 
\begin{equation}
\cos (a\varphi-\alpha )>0 \ \ \ \ {\rm for}\ \ \  0\leq \varphi<2\pi, 
\label{5.1}
\end{equation} 
$\phi_0^-(u_a(\alpha ))$ can be normalizable. 
The relation can be fulfilled by the suitable 
choices of the parameters such that $0<a<1/2$ and 
$-(1/2-2a)\pi<\alpha<\pi/2$. 
There are, of course, different choices, when we take the different 
solutions from 
$\phi_0^\pm(u_a(\alpha))$. 
It is very hard to answer the question 
whether the choice of the solutions is physically meaningful or not.  
Such solutions, however, possibly have some meanings 
in phenomena limited in very special regions, provided that 
the solutions are  used only in the limited regions and 
smoothly connected to other functions defined outside the regions. 
In fact the solutions are used for constructing 
the vortices from the plane-wave solutions in three-dimensional space. 
(See the argument of section 5.) 

Note also here that the solutions $\phi_0^\pm(u_a(\alpha ))=M_a 
e^{\pm k_au_a(\alpha )}$ 
have no current because  they can be 
taken as real. 
The higher polynomial solutions with $n\geq 2$ of (25) or (26) 
can, however, have currents because they are generally complex. 
This means that we have a possibility for producing vortices from these 
solutions even if they will appear only in very limited regions. 

\vskip10pt

\section{Remarks on non-zero energy solutions }
\label{sect.6.0} 

We shall briefly discuss the equation for non-zero energy given by 
\eqref{2.1.6}
$$
[-{\hbar^2 \over 2m}{\boldsymbol{\vartriangle}}_a-
  a^{-2}{\cal E}\ \rho_a^{2(1-a) / a}]\ \psi(u_a,v_a)=
    g_a \ \psi(u_a,v_a). \ \ \ \ \ \ \ \ \ 
$$ 
As noted in section 2, this equation can be read as the equation 
for determining the strength of the coupling constant $g_a$ of 
the original potential $V_a=-a^{2}g_a\rho^{2(a-1)} $ 
for the given energy ${\cal E}$. 
We shall, however, discuss it from a little different standpoint. 
If we can solve the eigenvalue problem for the potential of 
$-a^{-2}{\cal E}\ \rho_a^{2(1-a) / a} $, 
we can obtain the eigenvalues of the original equation 
$$
[-{\hbar^2 \over 2m}{\boldsymbol{\vartriangle}}-a^{2}g_a\rho^{2(a-1)}]\ \ \psi(x,y)=
    {\cal E}\  \psi(x,y).\ \ \ \ \ \ \ \ \ \ \ \ \ 
$$ 
Let us show one example for $a=1/2$, where the original potential 
is written by 
\begin{equation}
V_{1/2}(\rho)=-{1 \over 4}g_{1/2}{1 \over \rho} \ \ \ \ 
{\rm for}\  g_{1/2}>0. 
\label{6.2.1} 
\end{equation} 
For real and negative eigenvalues (${\cal E}<0$) the equation 
\eqref{2.1.6} can be understood as the two-dimensional harmonic 
oscillator with the spring constant 
$k=8|{\cal E}|$. 
The eigenvalues of the two-dimensional harmonic oscillator are 
well known as 
\begin{equation}
E_{n_x,n_y}=(n_x+n_y+1)\hbar \omega , 
\label{6.2.2} 
\end{equation} 
where $n_x$ and $n_y$ are zero or positive integers, and 
$\omega=2\sqrt{2|{\cal E}|/m}$. 
Thus we have the relation 
\begin{equation}
g_{1/2}=E_{n_x,n_y}. 
\label{6.2.3} 
\end{equation} 
From this relation we obtain the eigenvalue ${\cal E}$ as 
\begin{equation}
{\cal E}=-{ m g_{1/2}^2 \over 8(2N+1)^2\hbar^2}, 
\label{6.2.4} 
\end{equation} 
with $N=(n_x+n_y)/2$. 
We can directly confirm the eigenvalues by solving the 
original equation for the solutions 
$\psi(x,y)=R(\rho)e^{i l_z\varphi}$ ($l_z=$integers) 
that correspond to the symmetric 
solutions of the harmonic oscillator described by 
$n_x=n_y$. 
We see that, provided that one of the eigenvalue problems 
can be solved, we can also obtain the eigenvalues of the other 
equation. 
It is interesting that harmonic oscillator ($\rho^2$) and Coulomb type 
($\rho^{-1}$) are mapped each other by the conformal mapping 
and 
there is a relation between 
the energy eigenvalues of the two potentials in two dimensions. 

\vskip10pt

\section{Concluding remarks}
\label{sect.7.0}

We have shown that all Schr\"{o}dinger equations 
with the symmetric potentials of the type $V_a(\rho)$ 
in two dimensions can be reduced to 
the same equation with a constant potential 
for the zero energy eigenstates in terms of the conformal mappings and 
the states with the zero energy are in the infinite degeneracy. 
The degeneracy becomes not only the origin of the huge variety 
of vortex patterns but it will possibly be 
a interesting tool to investigate 
complicated problems of surface physics including boundaries 
as well. 
And the idea can be extended to phenomena in three dimensions. 
Particularly this scheme will become a powerful tool to vortex phenomena. 
Actually a vortex lattice solution has been found in this scheme~\cite{koba}. 
We may expect that hydrodynamical approach in quantum mechaincs 
presened here will open many interesting aspects in physics 
such as the investigation of vortex patterns 
[24-34]. 
We also note that the eigenstates satisfying the anyon boundaries 
will be applicable to problems in 
condensed matter physics such as vortex matters~\cite{blat,crab} 
and quantum Hall effects [45-49]. 

It should be noticed that some kinds of 
equations in hydrodynamics [38-41] 
are obtainable from the original 
eigenvalue equation~\eqref{2.1.1}   
by changing parameters such as $\hbar$ and mass $m$.  
This means that the conformal mappings~\eqref{2.1.2} are applicable 
to hydrodynamical problems in two dimensions 
and the infinite degeneracy can also take place. 
The analysis in terms of the functions obtained in this article 
will also become an interesting approach in many aspets of 
hydrodynamical problems. 
\vskip5pt

Finally we would like to note that the infinite degeneracy of 
the zero energy solutions brings infinite variety to many-body 
systems with a fixed energy, which possibly becomes an origin 
of a new entropy different 
from the Boltzmann entropy [42-44]. 
The new entropy has nothing to do with 
the determination of usual temperatures in thermal equilibrium but 
the new freedom stored in the new entropy can be released in 
thermal non-equilibrium~\cite{ks6}. 
These considerations will also bring a very new aspect in statistical 
mechanics extended from Hilbert spaces 
to Gel'fand triplets~\cite{ks5,ks7}.


\end{document}